\documentclass[aps,prd,nofootinbib,nobibnotes,notitlepage,twocolumn]{revtex4-2}
\usepackage{graphicx}
\usepackage{amssymb}
\usepackage{amsmath} 
\usepackage{color}
\usepackage{float}
\usepackage{ulem}
\usepackage{accents}

\usepackage{amsfonts}
\usepackage[colorlinks=true,
pdfstartview=FitV,linkcolor=blue,
citecolor=blue,urlcolor=blue,breaklinks=true]{hyperref}
\usepackage{array}
\usepackage{float}
\usepackage{placeins}
\usepackage[dvipsnames]{xcolor}
\usepackage{csquotes}
\usepackage[makeroom]{cancel}
\usepackage{units}
\usepackage{fixmath}
\usepackage{bigints}
\usepackage{amssymb}
\usepackage{systeme}
\newcolumntype{C}[1]{>{\centering\arraybackslash}m{#1}}
\renewcommand{\eqref}[1]{\mbox{Eq.~(\ref{#1})}}

\definecolor{ForestGreen}{rgb}{0.13,0.55,0.13}

\makeatletter
\renewcommand*\l@section{\@dottedtocline{1}{0em}{1.5em}}
\renewcommand*\l@subsection{\@dottedtocline{1}{1.5em}{1.5em}}
\renewcommand*\l@subsubsection{\@dottedtocline{1}{3em}{1.5em}}
\makeatother

\begin{document}
	\title{Double rotatory power reversal, continuous Kerr angle, and enhanced reflectance in bi-isotropic media with anomalous Hall current}

\author{Alex Q. Costa$^a$}
\email{costa.alex@discente.ufma.br, prof.costalex@gmail.com}
	\author{Pedro D. S. Silva$^{a,b}$}
	\email{pedro.dss@ufma.br, pdiego.10@hotmail.com}
	\author{Manoel M. Ferreira Jr.$^{a, c}$}
	\email{manojr.ufma@gmail.com, manoel.messias@ufma.br}
		\affiliation{$^a$Programa de P\'{o}s-gradua\c{c}\~{a}o em F\'{i}sica, Universidade Federal do Maranh\~{a}o, Campus
		Universit\'{a}rio do Bacanga, S\~{a}o Lu\'is, Maranh\~ao, 65080-805, Brazil}
		\affiliation{$^b$Coordena\c{c}\~ao do Curso de Ci\^encias Naturais - F\'isica, Universidade Federal do Maranh\~ao, Campus de Bacabal, Bacabal, Maranh\~ao, 65700-000, Brazil}
	\affiliation{$^c$Departamento de F\'{i}sica, Universidade Federal do Maranh\~{a}o, Campus
		Universit\'{a}rio do Bacanga, S\~{a}o Lu\'is, Maranh\~ao, 65080-805, Brazil}

	\begin{abstract}

We investigate the optical properties of bi-isotropic materials under the anomalous Hall effect (AHE) of axion electrodynamics. Four refractive indices associated with circularly polarized waves are achieved, implying circular birefringence with rotatory power endowed with double sign reversal, an exotic optical signature for chiral dielectrics. The Kerr rotation and ellipticity are analyzed, with an unusual observation of a giant rotation angle deprived of discontinuity. Anomalous enhanced reflectance (R greater than unity) is also reported, associated with negative refraction stemming from the anomalous transport properties. These effects constitute the singular optical signature of a nonequilibrium bi-isotropic medium with the AHE.

\end{abstract}

\maketitle
	
\section{Introduction} 

The optical characterization of chiral media constitutes a powerful technique for matter characterization. Birefringence appears as a key feature of optically active media, being measured by the rotatory power (RP) \cite{Condon}, which finds application in a broad set of scenarios, including  bi-isotropic matter \cite{Sihvola,Qiu2}, organic compounds \cite{Xing-Liu}, chiral metamaterials \cite{Zhang}, graphene based devices \cite{Amin}, sensors and filters \cite{Amin-Khan}, chiral metasurfaces \cite{Plum}, polarization rotators \cite{Zhanni-Wu}, dielectric and quasi-planar nanostructures \cite{Alexander-Zhu, Makoto}. The dispersive RP can be anomalous if it undergoes reversion with the frequency \cite{Newnham}. The Kerr rotation angles provide magneto-optical signatures of reflected light \cite{Shinagawa,Sato}. The magneto-optic Kerr effect (MOKE) and magneto-optical effects in complex materials are broadly employed to probe properties of topological insulators \cite{Ohnoutek}, new graphene composites \cite{Shimano} and to examine giant Kerr rotation in Weyl semimetals \cite{Sonowal} and time-reversal broken systems  \cite{Tse-MacDonald-PRL}.

Nonconventional phenomena, such as the Chiral Magnetic Effect (CME), where an electric current is generated along an applied magnetic field due to an imbalance in the number density of chiral fermions with opposite handedness \cite{Kharzeev1, Fukushima}, has played an important role in Weyl metals \cite{Burkov} and semimetals \cite{Gobar,Kaushik1}.  Another relevant quantum-anomaly induced transport phenomenon is the anomalous Hall effect (AHE), due to the separation of Weyl nodes in momentum for right- and left-handed fermions. The AHE and the CME can be effectively described by including the axion term \cite{Wilczek,Sekine, KDeng, Barnes}, $\mathcal{L}=\theta(\mathbf{E}\cdot\mathbf{B})$, in the classical Maxwell Lagrangian in continuous matter, where the anomalous Hall current takes place, $\nabla\theta\times\mathbf{E}$. For the special case of a nondynamical axion field, $\partial_{t}\theta=cte$, $\nabla\theta=cte$, one recovers the  Maxwell-Carroll-Field-Jackiw (MCFJ) electrodynamics \cite{CFJ} in continuous matter \cite{Qiu1,Pedro2}. For a time-independent axion field,  $\partial_{t}\theta=0$, the Ampere's law reads $\nabla\times\mathbf{H}-\partial_{t}\mathbf{D}=\mathbf{J}+\mathbf{b}\times\mathbf{E}$,
with $\mathbf{b}=\nabla\theta$ standing for the axion field gradient.  The axion electrodynamics effectively describes relevant aspects of Weyl semimetals \cite{Marco,Armitage}, optical properties of exotic metamaterials \cite{Barredo},  axion dielectrics \cite{Chang1,PedroPRB2024A, Martin-Ruiz}, connections with the London equation and Weyl semimetals \cite{Stalhammer,Shyta}, Cherenkov radiation \cite{Cherenkov},  applications in ultrafast magnetism \cite{Barredo2}, photonics of new chiral materials \cite{Barredo3},
and optical reflection properties at the surface of an axion dielectric \cite{PRB-Pedro-Ronald}. Wave propagation in bi-isotropic media endowed with the axion magnetic current term was classically addressed as well, yielding dispersive birefringence and rotatory power endowed with sign reversal \cite{PedroPRB02022}, a typical property of rotating plasmas \cite{Gueroult2}, chiral plasmas \cite{Filipe1}, graphene systems \cite{Poumirol}, and particular Weyl semimetals \cite{Day-Nandy}. Investigations on reflection for normal incidence at a Weyl semimetal surface have yielded anomalous reflectance (greater than unity) due to chiral magnetic instabilities associated with CME and AHE axion terms \cite{Nishida1,Nishida,Burkov2}. Moreover, nonreciprocal thermal radiation emission in Weyl semimetals (with AHE) has shown negative emissivity at low frequencies, deviating from Planck's law \cite{Konabe}. Kerr rotation and ellipticity angles \cite{Trepanier} depending on the frequency were reported in Weyl semimetals  \cite{Ghosh}, where it can be used to design circular polarizers or optical isolators \cite{Cote-Trepanier, Cheng-Guo}. 

In this paper, we investigate the optical characterization of bi-isotropic media endowed with the AHE, including circular birefringence, Kerr rotation, Kerr ellipticity, and anomalous reflection, achieving new remarkable features that compose a specific optical signature of such a system. Throughout this work, we use natural units.

\section{Dispersion relations and refractive indices} 

In the plane wave ansatz, one writes the Maxwell equations modified by the AHE term,
\label{maxwell-equations-plane-wave-1}
	\begin{align}
		\mathbf{k}\cdot\mathbf{D}   =  -i \mathbf{b} \cdot \mathbf{B} , \quad \quad \mathbf{k}\times\mathbf{H}   =-\omega\mathbf{D} + i	\,	\mathbf{b}\times\mathbf{E},   \label{maxwell-equations-1} 
\end{align}
which must be considered with the linear bi-isotropic constitutive relations
	\begin{align}
			\mathbf{D} =\epsilon\mathbf{E}+\alpha\mathbf{B}, \quad 
		\mathbf{H}=\dfrac{1}{\mu}\mathbf{B}+\beta\mathbf{E},
		\label{CRBi1}
	\end{align}
where $\alpha$ and $\beta$ are complex coefficients that fulfill the condition $\beta^*=-\alpha$ to assure energy conservation. Manipulating the Maxwell equations  and \eqref{CRBi1}, one finds the equation $M_{ij}E^j=0$,
with $M_{ij}=n^2\delta_{ij}-n_{i}n_{j}-\mu\bar{\epsilon}_{ij}$, and the effective electric permittivity tensor,
\begin{equation}
\bar{\epsilon}_{ij}\left(\omega\right)=\epsilon\delta_{ij}+\dfrac{1}{\omega} (\alpha + \beta)  \epsilon_{ilj}k^{l}-\frac{i}{\omega}\epsilon_{ilj}b^{l}, \label{permittivity1}
\end{equation}
which includes the magnetoelectric and the AHE contributions. Here, $n_{i}$ is a refractive index component, $\mathbf{n}={\mathbf{k}}/{\omega}$.
Solving $\det[M_{ij}]=0$ provides the refractive indices 
	\begin{align}
		\label{RCPLCP} 
		n_{1,\pm}   &=\mu\alpha^{\prime\prime}\pm Q_{-}, \quad
	n_{2,\pm}     =-\mu\alpha^{\prime\prime}\pm Q_{+},
	\end{align}
where  $\mathbf{n}$ is considered parallel to the vector $\mathbf{b}$, 
\begin{equation}
		Q_{\pm}=\sqrt{ \mu \epsilon +\mu^2\alpha^{\prime\prime2} \pm \mu b/\omega}, \label{Qpm} 
\end{equation}
and $\alpha+\beta=2i\alpha^{\prime\prime}$, with $\mathrm{Im}[\alpha]=\alpha''$. The propagation modes for waves along the z-axis are described by the polarization vectors  $\hat{\bf{e}}_{\mp} = \hat{\bf{x}} \mp i \hat{\bf{y}}$  -- right-handed circularly polarized (RCP) and left-handed circularly polarized (LCP) waves, respectively. The two indices $n_{1,\pm}$ are associated with the RCP polarization vector ($\hat{\bf{e}}_{-}$).  While $n_{1,+}$ designates forward RCP propagation for any frequency, the negative index $n_{1,-}$ corresponds to forward RCP propagation in the frequency range $\omega < \omega_{R}$ and backward LCP propagation for $\omega > \omega_{R}$, where $\omega_{R}=b/\epsilon$. The indices $n_{2, \pm}$ are associated with the LCP vector ($\hat{\bf{e}}_{+}$), with $n_{2,+}$ indicating forward LCP propagation and $n_{2,-}$ a backward RCP wave. The four solutions in \eqref{RCPLCP} are distinct from one another. Along with that, one observes that the propagation of RCP forward and backward waves are associated with two distinct refractive indices (when considering a global minus sign, due to $\pm \mu b/\omega$ factor in $Q_{\pm}$). The same is true for the forward and backward LCP waves. This is a manifestation of nonreciprocity (stemming from the AHE term).

\begin{figure}[H]
\centering
	\includegraphics[scale=0.6]{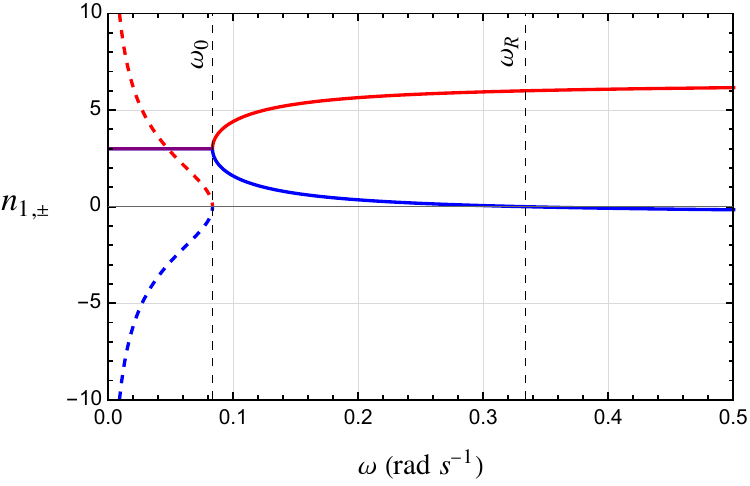}
	\caption{Refractive indices $n_{1,\pm}$ of \eqref{RCPLCP}. The red (blue) lines represent $n_{1,+}$ ($n_{1,-}$). The solid (dashed) line indicates the real (imaginary) part of the refractive indices. The solid purple line represents the real parts of $n_{1,+}$ and $n_{1,-}$,  lying on top of each other. Here, we have used $\mu=1, \epsilon=3, \alpha^{\prime\prime}=3$ and $b=1$ $\mathrm{s}^{-1}$.} \label{RCP_plots}
\end{figure}

The indices $n_{1,\pm}$ are complex in the frequency range $\omega<\omega_{0}$, with
\begin{equation}
	\omega_0={ b}/({\epsilon+\alpha^{\prime\prime2}\mu}) , 	\label{omegazero1}
\end{equation}
where $0<\omega<\omega_{0}$ is the absorption zone (for positive definite electromagnetic parameters). In Fig.~\ref{RCP_plots}, we illustrate the frequency behavior of the indices $n_{1,\pm}$ of \eqref{RCPLCP}. For $\omega>\omega_{0}$, the indices become real, being $n_{1,+}$ (red line) positive and monotonically increasing (normal dispersion), while $n_{1,-}$ (blue line) is progressively decreasing (anomalous dispersion).
\begin{figure}[H]
\centering
	\includegraphics[scale=0.6]{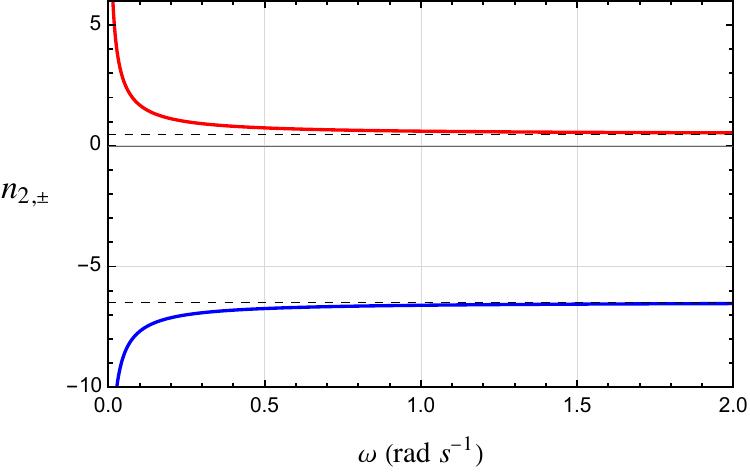}
	\caption{Refractive indices $n_{2,\pm}$ of \eqref{RCPLCP}. The red (blue) line indicates the positive (negative) refractive index ($n_{2,\pm}$), respectively. The horizontal dashed lines are given by Eq. (\ref{Asymp1}). Here, we have used $\mu=1, \epsilon=3, \alpha^{\prime\prime}=3$ and $b=1$ $\mathrm{s}^{-1}$.}
	\label{LCP_plots}
\end{figure}	

  Furthermore, the index $n_{1,-}$ becomes negative for $\omega>\omega_R $, with $\omega_R=b/\epsilon$, as shown in Fig.~\ref{RCP_plots}. Also, the refractive indices $n_{2,\pm}$ are both real, being $n_{2,-}$ negative for any frequency, while $n_{2,+}$ is always positive, as shown in Fig.~\ref{LCP_plots}.  In the high-frequency regime, the indices $n_{1,\pm}$ and $n_{2, \pm}$ tend to positive and negative asymptotic values,
  \begin{align}
   	\left. n_{1,\pm} \right|_{\mathrm{high}} &= \mu \alpha^{\prime \prime} \pm \sqrt{ \mu \epsilon + \mu^{2} \alpha^{\prime \prime 2}}, \label{n-RCP-high-frequency-1} \\
  	\label{Asymp1}
  	\left.	n_{2,\pm} \right|_{\mathrm{high}}   &=-\mu\alpha^{\prime\prime}\pm\sqrt{\mu\epsilon+\mu^{2}%
  		\alpha^{\prime\prime2}}, 
  \end{align}
  which represent the frequency-independent indices of a bi-isotropic dielectric.

\section{Birefringence and Rotatory Power}

The circular birefringence is measured in terms of the rotatory power,
\begin{equation}
	\delta_{RP}=-\dfrac{\omega}{2} \left[\operatorname{Re}(n_{LCP})\ - \operatorname{Re}(n_{RCP})\right]  ,
\end{equation}
for waves propagating in the same direction. The fact that we have four refractive indices, associated with forward and backwards waves, allows us to write the following expressions for rotatory power:
	\begin{align}
		\delta_{+,\pm} &= \frac{\omega}{2}\operatorname{Re} \left(
		 2\mu\alpha^{\prime\prime} - Q_{+} \pm Q_{-} \right), \label{RP+pm} \\
\delta_{-,-} &= \frac{\omega}{2}\operatorname{Re} \left(
2\mu\alpha^{\prime\prime} +  Q_{+}  - Q_{-}  \right), \label{RP-pm}
\end{align}
where the first subscript ($\pm$) stands for $n_{2, \pm}$, and the second one for $n_{1, \pm}$. Notice that one can not write $\delta_{-, +}$ since it is related to a backward RCP wave with $n_{2-}$ and a forward RCP wave with $n_{1,+}$. Furthermore, $\delta_{+, -}$ holds in the frequency range $\omega < \omega_{R}$, where the mode associated with $n_{1,-}$ corresponds to forward propagation of an RCP wave.

\begin{figure}[h]
	\centering\includegraphics[scale=0.4]{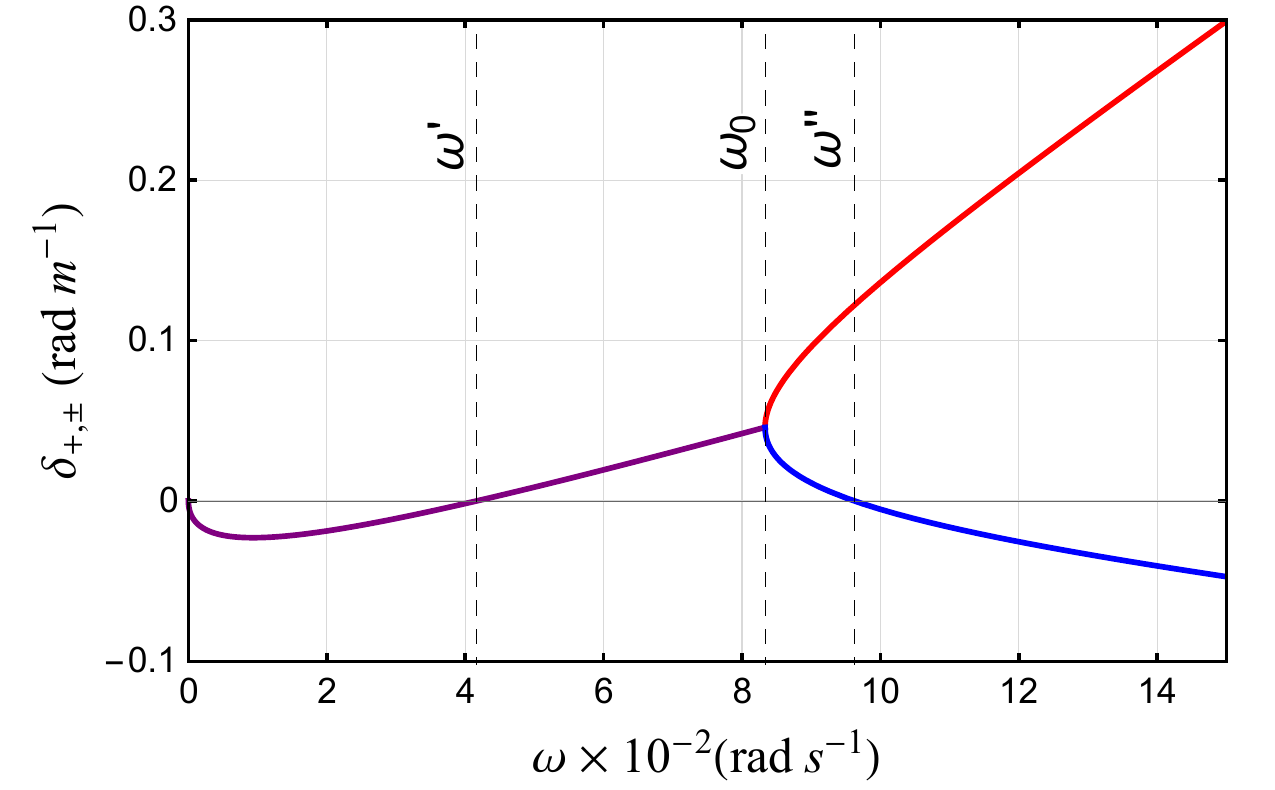}
\caption{Rotatory power $\delta_{+,\pm}$ of \eqref{RP+pm}, constructed using the two refractive indices for the RCP wave and $n_{2,+}$.  The RP $\delta_{+,+}$ is depicted by the purple-red line, while $\delta_{+,-}$ is demarcated by the purple-blue line, endowed with double sign reversal. Here, we have used $\mu=1, \epsilon=\alpha^{\prime\prime}=3$, and $b=1$ $\mathrm{s}^{-1}$.} 
\label{RP1}
\end{figure}

The behavior of the rotatory power $\delta_{+,\pm}$ of Eq.~(\ref{RP+pm}) as a function of the frequency is observed in Fig.~\ref{RP1}, being both equal in the range $0<\omega<\omega_0$ (see the purple line), where they only receive contribution\footnote{One notes that $Q_{-}$ contributes to the RP only when it is real, that is, for {$\omega>\omega_0$}, with $\omega_0$ given by Eq. (\ref{omegazero1}).} of the first root ($Q_{+}$). In this frequency range, both RPs  $\delta_{+,\pm}$ possess sign reversal at $\omega=\omega'$, with $\omega'$ given by $\omega^{\prime}={b}/({3\mu\alpha^{\prime\prime2}-\epsilon})$. For $\omega>\omega_0$, the RPs begin to receive the contribution of the second root, resulting in the splitting starting at $\omega=\omega_0$, from which $\delta_{+,+}$ exhibits an increasing behavior (see the red line), while $\delta_{+,-}$ has decreasing magnitude (see the blue line).  Along with that, the RP $\delta_{+,-}$ exhibits a second (additional) sign reversal at the frequency, $\omega^{\prime\prime}={b}/({2\alpha^{\prime\prime}\sqrt{\mu\epsilon}})$,
with $\omega^{\prime\prime}>\omega_0$. This double sign inversion is an effect of both the magnetoelectric and AHE parameters, a very unusual feature that may work as an optical signature of this chiral system. A single RP reversion was reported for the case of a bi-isotropic dielectric in the presence of an isotropic axion magnetic current \cite{PedroPRB02022}.
For high frequencies, where the term $b/\omega$ becomes negligible, the coefficients $\delta_{+,+}$ and $\delta_{+,-}$ recover a linear behavior with a positive (negative) slope.

\section{Kerr rotation and Kerr ellipticity}

Considering incident light stemming from medium 1 (a usual dielectric described by permittivity $\epsilon_1$ and permeability $\mu_1$) and reflecting on the surface of medium 2 [a chiral dielectric described by the parameters $\epsilon$, $\mu$, the axion vector $\mathbf{b}$, and the refractive indices of \eqref{RCPLCP}], we obtain two Fresnel coefficients for each circularly polarized mode associated with $n_{1, \pm}$ and $n_{2, \pm}$ at normal incidence, namely
\begin{subequations}
\label{general-Fresnel-coefficients-1}
	\begin{align}
		\label{eqfresnelA}
		r_{2,\pm}   =\frac{\mu_{1}n_{2,\pm}-\mu n_{1}}{\mu_{1}n_{2,\pm}+\mu  n_{1}}, \\
		r_{1,\pm}   =\frac{\mu_{1}n_{1,\pm}-\mu n_{1}}{\mu_{1}n_{1,\pm}+\mu n_{1}}.
		 \label{eqfresnelB}
	\end{align} 
\end{subequations} 
The latter can be used to define the Kerr angle expressions \cite{PRB-Pedro-Ronald}
\begin{align}
\Delta_{+, \pm} &= \left(  \frac{r_{2,+}-r_{1,\pm}}{r_{2,+}+r_{1,\pm}}\right), \label{complex-Kerr-angle-parameter-1} \\
\Delta_{-, -} &= \left(  \frac{r_{2,-}-r_{1,-}}{r_{2,-}+r_{1,-}}\right), \label{complex-Kerr-angle-parameter-1-extra-1} 
\end{align}
where the first (second) $\pm$ subscript is related to $n_{2, \pm}$ ($n_{1, \pm}$). The quantity $\Delta_{+,-}$ holds only in the frequency range $\omega < \omega_{R}$. The elements $\Delta_{+,\pm}$ and $\Delta_{-, -}$ are generally complex, 
\begin{align}
	\Delta_{+,\pm}=\Delta_{+,\pm}^{\prime}+i\Delta_{+,\pm}^{\prime\prime},
\end{align}
with $\Delta_{+,\pm}^{\prime}=\operatorname{Re}[\Delta_{+,\pm}]$ and $\Delta_{+,\pm}^{\prime\prime}=\operatorname{Im}[\Delta_{+,\pm}]$. For $\Delta_{-,-}$, one writes analogous expressions.
The real and imaginary parts provide the Kerr angles (ellipticity and rotation) \cite{PRB-Pedro-Ronald, Sato}, respectively, 
\begin{align}
	\tan ( \,  2\theta_{K}^{+,\pm} \, ) &=-\frac{2\Delta_{+,\pm}^{\prime\prime}
	}{1-\left\vert \Delta_{+,\pm}\right\vert ^{2}}, \label{thetaK1} \\
	\sin ( \, 2\eta_{K}^{+,\pm} \,) &=\frac{2\Delta_{+,\pm}^{\prime}
	}{1+\left\vert \Delta_{+,\pm}\right\vert ^{2}}.
	\label{KE}
\end{align}

Recalling the refractive indices in \eqref{RCPLCP}, we note that $n_{2, +}$ and $n_{1, +}$ are connected to the usual refraction (for $\omega< \omega_{R}$). On the other hand, the indices $n_{2, -}$ and $n_{1, -}$ (for $\omega > \omega_{R}$) are related to negative refraction. Thus, the element $\Delta_{+, +}$ provides the Kerr angles for the usual refraction case,  $\Delta_{-, -}$ yields the Kerr angles for the pure negative refraction scenario for ($\omega > \omega_{R}$).

\begin{figure}[h]
\centering
	\includegraphics[scale=0.6]{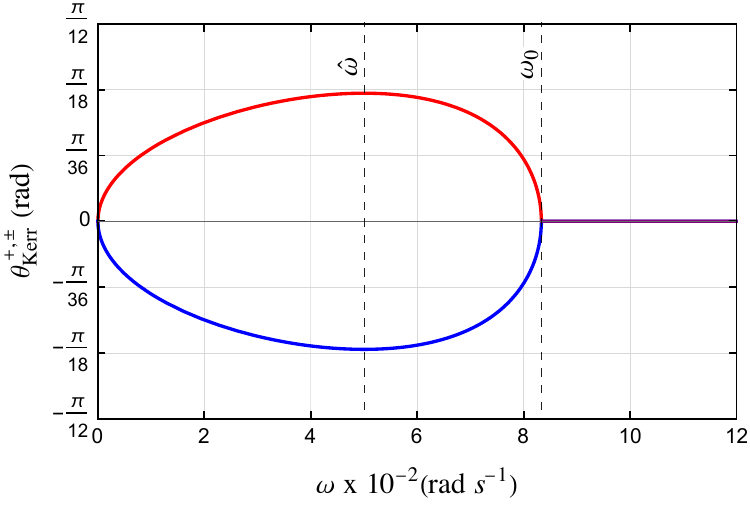}
	\caption{Kerr rotation $\theta_{K}^{+,\pm}$ of \eqref{thetaK1}. The red line shows the behavior of $\theta_{K}^{+,+}$,  composed by the indices $n_{2,+}$ and $n_{1,+}$. The blue line depicts $\theta_{K}^{+,-}$, constituted with $n_{2,+}$ and $n_{1,-}$.  The purple line is defined for the region where both $\theta_{K}^{+,\pm}$ are null. Here, we have used $n_{1}=1$, $\mu_{1}=1$ $\mu=1$, $\epsilon=3, \alpha^{\prime\prime}=3$, and $b=1$ $\mathrm{s}^{-1}$.} \label{Kerr_rotation}
\end{figure}

As previously discussed, absorption occurs only in the range $\omega<\omega_0$, being associated with $n_{1,\pm}$ refractive indices. Thus, only in this frequency interval, one finds $\Delta^{\prime\prime}\neq 0$, providing a non-null Kerr rotation for the reflected wave. In the following subsection, we examine the Kerr rotation angles for $\Delta_{+, \pm}$ in terms of frequency.

The general behaviors of $\theta_{K}^{+,\pm}$ in terms of the frequency are illustrated in Fig.~\ref{Kerr_rotation}. We observe that the factors $\theta_{K}^{+, +}$ and $\theta^{+,-}_{K}$ are positive and negative, respectively, meaning that the polarization ellipse of the reflected wave is rotated in the counterclockwise (clockwise) direction relative to $x$-axis of polarization basis \cite{Collett}.   

The giant continuous Kerr rotation angle of Fig.~\ref{Kerr_rotation} can be close to the magnitude $\pi/18$ and represents a remarkable peculiar characteristic of this bi-isotropic system endowed with AHE term, in contrast to the usual discontinuous Kerr rotation angles that occur in media ruled by the axion electrodynamics \cite{Sonowal}. For the dielectric system considered here, such a discontinuity takes place only when
\begin{align}
  n_{1} > \mu_{1} \alpha^{\prime \prime},  \label{condition-for-divergence-Kerr-rotation-1}
\end{align}
for which the squared magnitude of coefficient $\Delta_{+,+}$ becomes equal to unity, leading to a divergence for the  Kerr rotation angle tangent ($\theta_{K}=\pi/4$). The latter manifests itself as an abrupt sign change (discontinuity) in $\theta_{K}^{+, +}$ at the frequency $\tilde{\omega}$ given by $\tilde{\omega} = {\mu_{1}^{2} b}/({\mu_{1}^{2} \epsilon + \mu n_{1}^{2}})$.

\begin{figure}[h]
	\includegraphics[scale=0.6]{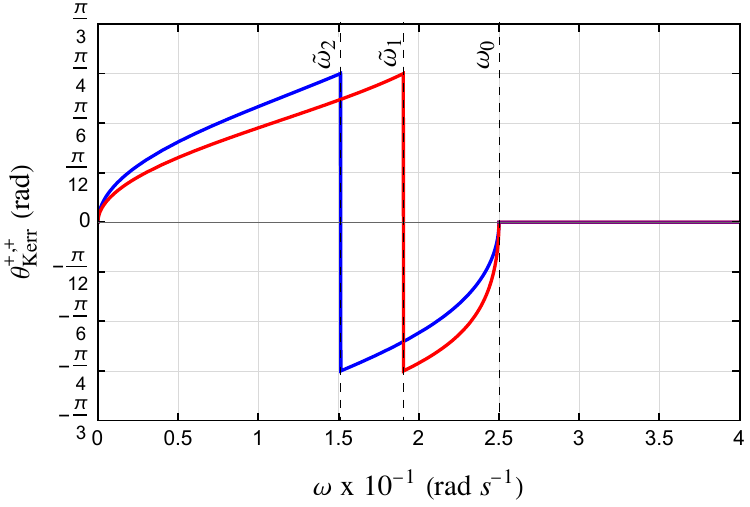}
	\caption{Kerr rotation $\theta_{K}^{+,+}$ of \eqref{thetaK1} under the condition (\ref{condition-for-divergence-Kerr-rotation-1}). The red and blue lines show $\theta_{K}^{+,+}$ (obtained by using $n_{2,+}$ and $n_{1,+}$) for $n_{1}=1.5$ and $n_{1}=1.9$, respectively. The purple line indicates $\theta_{K}^{+,+} =0$. Here, we have used $\mu_{1}=1$ $\mu=1$, $\epsilon=3, \alpha^{\prime\prime}=1$, $b=1$ $\mathrm{s}^{-1}$, $n_{1}=1.5$ (red), and $n_{1}=1.9$ (blue). } \label{plot-Kerr-rotation-with-divergence}
\end{figure}

Figure \ref{plot-Kerr-rotation-with-divergence} illustrates the frequency behavior of $\theta_{Kerr}^{+, +}$ under condition (\ref{condition-for-divergence-Kerr-rotation-1}). Notice that at the frequencies $\tilde{\omega}_{1}$ and $\tilde{\omega}_{2}$ for the red (blue) line, respectively, the rotation of the polarization ellipse changes abruptly, indicating a reversal in the ellipse polarization rotation.  The other Kerr rotation angles, $\theta_{K}^{-, -}$, which are obtained by using $n_{2, -}$, can be easily found by implementing $\Delta_{-, -} = \left. \Delta_{+, +} \right|_{Q_{+} \rightarrow -Q_{+}}$, and will not be addressed here.

\begin{figure}[h]
\centering
	\includegraphics[scale=0.6]{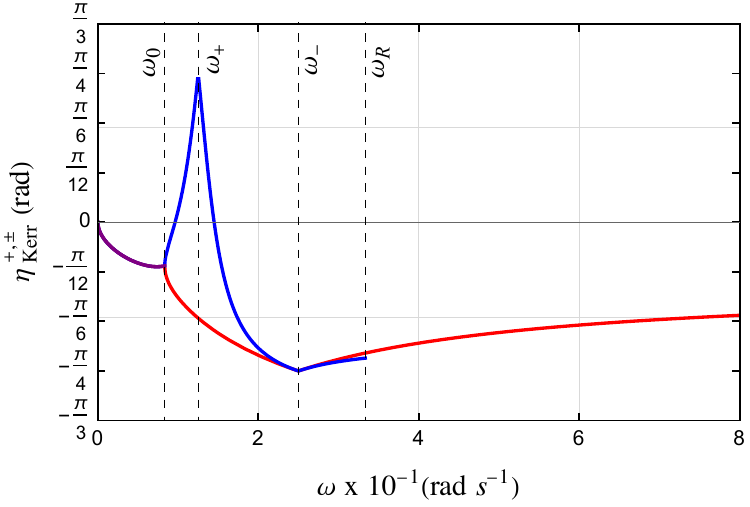}
	\caption{Kerr ellipticity $\eta_{K}^{+, \pm}$ of \eqref{KE}. The red line shows $\eta^{+,+}_{K}$, while the blue curve depicts $\eta^{+,-}_{K}$. The purple line indicates the region both $\eta_{K}^{+,\pm}$ are equal to each other. Here, we have used $n_{1}=1$, $\mu_{1}=1$, $\mu=1, \epsilon=3, \alpha^{\prime\prime}=3$, and $b=1$ $\mathrm{s}^{-1}$. When $\eta_{K}=0$, the reflected wave is linearly polarized. The extreme values ($\pm \pi/4$) are associated with left- and right-handed circularly polarized waves.} \label{plot-Kerr-ellipticity-angles}
\end{figure}

\section{Kerr ellipticity angles $\eta_{K}^{+, \pm}$}

The Kerr ellipticities, given by \eqref{KE},  are non-null for any frequency since the refractive indices (\ref{RCPLCP}) possess at least one real piece. For $0<\omega<\omega_0$, it holds $\operatorname{Re}[n_{1,\pm}]=\mu \alpha^{\prime\prime}$, so that the initial possibilities of Eq. (\ref{KE}) are reduced to only one. For the index $n_{2,+}$, we define $\eta_{K}^{+,\pm}$, whose behavior in terms of frequency is illustrated in Fig.~\ref{plot-Kerr-ellipticity-angles}. For $\omega<\omega_{0}$, one has $\eta^{+, +}_{K} = \eta^{+, -}_{K}$, as already explained, behavior represented by the purple line in Fig.~\ref{plot-Kerr-ellipticity-angles}. In the range $\omega>\omega_0$, the indices $n_{1, \pm}$ are real, assuming two different values, therefore making $\eta^{+,+}_{K} \neq \eta^{+,-}_{K}$, splitting the Kerr ellipticity into two distinct lines (red and blue, Fig.~\ref{plot-Kerr-ellipticity-angles}). In the frequencies $\omega_{\pm}$, given by
\begin{align}
\omega_{\pm} = \frac{ \mu_{1} b}{2 \mu \alpha^{\prime \prime} \sqrt{\mu_{1}\epsilon_{1}} \pm ( \mu_{1} \epsilon -\mu \epsilon_{1})}. 
\end{align}
the ellipticity is $\pm \pi/4$, respectively, where the wave is circularly polarized. As mentioned previously, $\Delta_{+,-}$ is only defined for $\omega < \omega_{R}$, being thus important to ensure that the frequencies $\omega_{+}$ and $\omega_{-}$ are within this range\footnote{The frequencies $\omega_{+}$ and $\omega_{-}$ are within the range $\omega < \omega_{R}$ when (i) $2\alpha^{\prime \prime} \sqrt{\mu_{1} \epsilon_{1}} > \epsilon_{1}$, for $\omega_{+} < \omega_{R}$; and (ii) $2\alpha^{\prime \prime} \sqrt{\mu_{1} \epsilon_{1}} > 2 \frac{\mu_{1}}{\mu}\epsilon - \epsilon_{1} $, for $\omega_{-} < \omega_{R}$; being it the case of Fig.~\ref{plot-Kerr-ellipticity-angles}.}, as shown in  Fig.~\ref{plot-Kerr-ellipticity-angles}.

 Moreover, the Kerr ellipticity $\eta^{+, -}$ presents a double reversion sign, indicating that the reflected wave can be turned into linearly polarized mode ($\eta^{+, -}=0$) at two distinct frequencies. This peculiar aspect seems to be in agreement with the double rotatory power reversal observed for $\delta_{+, -}$ in Fig.~\ref{RP1}.

\section{Reflectance and super reflectance effect}

\begin{figure}[h]
\centering
	\includegraphics[scale=0.6]{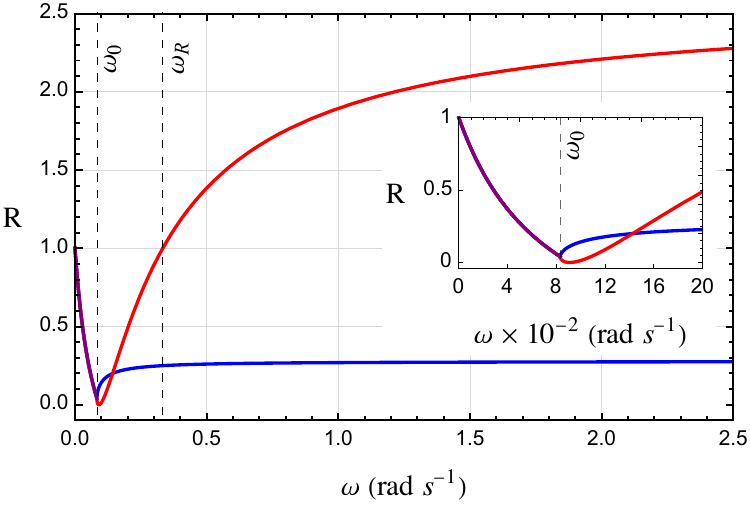}
	\caption{Reflection coefficients $R_{1}^{-}$ (red line) and $R_{1}^{+}$ (blue line). $R_{1}^{-}$ becomes greater than unit for $\omega>\omega_R$. The inset plot highlights the behavior near the frequency $\omega_{0}$, below which both $n_{1,\pm}$ have non-null imaginary parts. The solid purple line indicates that $R_{1}^{+}$ and $R_{1}^{-}$ are approximately equal to each other near the origin. Here, we have used $\mu_{1}=1, n_{1}=\sqrt{2}$, $\mu=1, \epsilon=3, \alpha^{\prime\prime}=3$, $b=1$ $\mathrm{s}^{-1}$, which implies $\omega_0=0.1$ and $\omega_R=1/3$ $\mathrm{s}^{-1}$.  } \label{Rmenos1}
\end{figure}

Recently, anomalous reflection ($R>1$), at normal incidence, has been reported as one key feature of Weyl semimetals in the presence of both axion terms ($b_0$ and $\mathbf{b}$) \cite{Nishida}. Here, we report that a similar effect can appear in bi-isotropic media modified by the AHE term. Starting from the Fresnel coefficients (\ref{eqfresnelA}) and (\ref{eqfresnelB}), one obtains the reflection coefficients,
\begin{subequations}
\label{reflectances-general-1}
	\begin{align}
R_{2}^{\pm} &=  \left|\dfrac{\mu_{1}\left(\mu\alpha^{\prime\prime} \mp Q_{+} \right)+\mu n_1}{\mu_{1}\left(\mu\alpha^{\prime\prime}\mp Q_{+}\right) - \mu n_1}\right|  ^2  ,\\
	R_{1}^{\pm} &=\left|\dfrac{\mu_{1}\left(\mu\alpha^{\prime\prime} \pm Q_{-}\right)-\mu n_1}{\mu_{1}\left(\mu\alpha^{\prime\prime}\pm Q_{-}\right)+\mu n_1}\right|^2,
\end{align}
\end{subequations}
where $Q_{\pm}$ are given in Eqs.~(\ref{Qpm}). We then address the two reflection amplitudes, $R_{1}^{+}$ and $R_{1}^{-}$,  stemming from $n_{1,+}$ (always positive for all frequencies) and $n_{1,-}$ (with negative real part for $\omega> \omega_R$). In Fig.~\ref{Rmenos1}, we illustrate reflectances that exhibit anomalous behaviors. The coefficient $R_{1}^{+}$ (blue line) already starts at unit, $R_{1}^{+}=1$, progressively diminishes with the frequency, reaching a mininimum near the frequency $\omega_{0}$, and then monotonically increases with frequency to its assymptotical value. On the other hand, $R_{1}^{-}$ (red line) initially decreases with frequency, reaches a null value near $\omega_0$, and then augments monotonically, becoming greater than unity for $\omega>\omega_R$, which is exactly the frequency whereupon the refractive index $n_{1,-}$ becomes negative. Thus, $R_{1}^{-}$ is endowed with super reflectance, being this property associated with negative refraction, a hallmark of metamaterials \cite{Valanju,Zhang}. The negative refraction of the index $n_{1}^{-}$ occurs as a consequence of an anomalous increasing wave magnitude that propagates along the direction $-\hat{z}$ \cite{Nishida}.

\section{Final remarks}

In this work, we investigated the optical properties of bi-isotropic materials endowed with the anomalous Hall current of axion electrodynamics. Four dispersive and distinct refractive indices were obtained, implying richer optical effects, such as rotatory power, Kerr rotation/ellipticity, and reflectance. Birefringence aspects were examined in terms of three distinct rotatory power coefficients. An RP endowed with double sign reversal was reported, being a remarkable characteristic of this particular dispersive bi-isotropic media with the AHE. The first signal reversal occurs for $\omega < \omega_{0}$, while the second one happens for $\omega > \omega_{0}$, the range in which the contribution of the full $n_{\text{1},-}$ index emerges, as shown in Fig.~\ref{RP1}.

We also considered an interface separating two media: dielectric medium 1 ($\epsilon_{1}$,$\mu_{1}$) and bi-isotropic medium 2 with AHE ($\epsilon$, $\mu$, $b$). For normal incidence, we have written three distinct Kerr rotation angles; see \eqref{thetaK1}. A continuous Kerr rotation angle (deprived of discontinuity) was found, reaching giant magnitudes as large as $\pi/18$, see Fig.~\ref{Kerr_rotation}, providing another key optical feature of bi-isotropic media with AHE. Such behavior occurs whenever the inequality (\ref{condition-for-divergence-Kerr-rotation-1}) does not hold.  In typical Weyl semimetals scenarios, the absence of the magnetoelectric parameter, $\alpha^{\prime \prime}=0$, assures the condition (\ref{condition-for-divergence-Kerr-rotation-1}) as granted, in such a way that the Kerr angle discontinuity is observed.  The Kerr ellipticity was also analyzed, with its dispersive behavior illustrated in Fig.~\ref{plot-Kerr-ellipticity-angles}. The reflected wave can exhibit linear polarization ($\eta=0$) at two distinct frequencies, left- and right-handed circular polarization ($\eta = \pm \pi/4$), and elliptical polarization (other values). Finally, the reflectance for normal incidence was also considered, starting from the Fresnel coefficients in \eqref{general-Fresnel-coefficients-1}. Four reflectances were carried out, with two of them depicted in Fig.~\ref{Rmenos1}, where an anomalous reflection behavior ($R>1$) becomes evident in the range $\omega > \omega_{R}$ for the red line ($n_{1,-}$), the exact region where this refractive index becomes negative. Super reflectance ($R >1$) was observed in a special class of materials (amplifying matter) under conditions of enhanced total reflection \cite{Cybulski-Silverman}, in which the refractive index of medium 1 is greater than that of medium 2 \cite{Plotz}. In our case, the coefficient $R>1$ happens as a consequence of the negative refractive index that occurs in certain frequency ranges, for which the refractive index of medium 2 becomes smaller than that of medium 1. In this sense, the phenomenon we have reported is analogous (at least in an effective way) to the enhanced reflection described in the 1970s. Furthermore, the increasing reflectance shown in Fig.~\ref{Rmenos1} is dispersive, differing from the non-dispersive scenario of Refs.~\cite{Cybulski-Silverman, Plotz}. A previously suggested explanation for $R>1$ attributes this effect to mechanisms occurring in medium 2 that supply energy to the reflected wave ~\cite{Plotz, Koester1966}.
In Weyl semimetals, with both chiral magnetic conductivity and AHE, the chiral imbalance drives the system out of equilibrium, and relaxation towards the equilibrium can supply the additional energy involved in the anomalous reflectance \cite{Nishida1, Nishida}.
	
The bi-isotropic media with AHE here examined also exhibits unstable electromagnetic modes, behaving very distinctly in three ranges of frequency:  (i) for $ 0 < \omega < \omega_{0}$, with $\omega_{0} = \mu b / (\mu \epsilon+ \mu^{2} \alpha^{\prime \prime 2})$, the refractive index $n_{1-}$ is complex, with its real piece positive and the imaginary part negative, describing a wave propagating with exponentially increasing amplitude in the z-axis positive direction. This causes an anomalous transmittance ($T>1$), implying $R+T>1$, which is a nonequilibrium effect; here 	$T = \left| \left\langle {\bf{S}}_{T} \right\rangle \cdot \hat{\bf{n}}\right|/ \left|  \left\langle {\bf{S}}_{I} \right\rangle \cdot \hat{\bf{n}} \right|$, with  $\mathbf{S}_{I}$ and $\mathbf{S}_{T}$ being the Poynting vector of incident and transmitted waves, respectively. (ii) For $\omega_{0} < \omega < \omega_{R}$, $\omega_{R} = b /\epsilon$,  the index $n_{1-}$ is real, positive and decreases with frequency (anomalous dispersion), so that $T$ continues to increase with frequency, keeping itself above 1 ($T>1$). (iii) For $\omega > \omega_{R}$, the index $n_{1-}$ becomes negative and the electromagnetic mode propagates backwards, yielding $T=0$ and $R>1$. This unstable behavior stems from both the bi-isotropic parameter $\alpha$ and the AHE, whose joint effect drives a nonequilibrium configuration. Thus, the additional energy for the anomalous reflected wave ($R >1$) or transmitted wave ($T >1$) can be interpreted as a consequence of the unstable electromagnetic waves in a nonequilibrium state. Such effects will be addressed in a forthcoming  manuscript \cite{Alex20025}.

Materials with bi-isotropic constitutive relations (or isotropic magnetoelectric tensors) can be found in the class of pyrochlores \cite{Sinisa}, such as Eu$_{2}$Ir$_{2}$O$_{7}$ and Nd$_{2}$Ir$_{2}$O$_{7}$. For both, anomalous Hall conductivity has been reported. See Ref.~\cite{Kentaro} and \cite{Xiaoran-Liu}. Other notable bi-isotropic materials include the chalcogenides Bi$_{2}$Se$_{3}$, Bi$_{2}$Te$_{3}$, and Sb$_{2}$Te$_{3}$ \cite{Lakhtakia-Mackay}. In particular, AHE has been observed in Bi$_{2}$Se$_{3}$, as reported in Refs.~\cite{Jing-Liu, Nan-Liu}. In principle,  bi-isotropic compounds in the presence of the AHE and non-equilibrium configurations gather the conditions for manifesting the effects discussed in this work, opening up exciting new avenues for further investigations on chiral media, as already done for Weyl semimetals in Refs.~\cite{Nishida1,Nishida,Burkov2}.

\section*{Acknowledgments} 

The authors thank FAPEMA, CNPq, and CAPES (Brazilian research agencies) for their invaluable financial support. M.M.F. is supported by FAPEMA APP-12151/22, CNPq/Produtividade 317048/2023-6 and CNPq/Universal/422527/2021-1. P.D.S.S. is grateful to FAPEMA APP-12151/22. Furthermore, we are indebted to CAPES/Finance Code 001 and FAPEMA/POS-GRAD- 04755/24.

\end{document}